\documentclass{article}

\usepackage{arxiv}

\usepackage[utf8]{inputenc} 
\usepackage[T1]{fontenc}    
\usepackage{url}            
\usepackage{booktabs}       
\usepackage{amsfonts}       
\usepackage{nicefrac}       
\usepackage{microtype}      
\usepackage{lipsum}
\usepackage{amsmath}

\title{Measurement error as a missing data problem}

\author{
  Ruth H.~Keogh \\
  Department of Medical Statistics\\
  London School of Hygiene and Tropical Medicine\\
  United Kingdom\\
  \texttt{ruth.keogh@lshtm.ac.uk} \\
   \And
Jonathan W.~Bartlett \\
  Department of Mathematical Sciences\\
  University of Bath\\
  United Kingdom \\
  \texttt{j.w.bartlett@bath.ac.uk} \\
}

\begin{document}
\maketitle

\begin{abstract}
 This article focuses on measurement error in covariates in regression analyses in which the aim is to estimate the association between one or more covariates and an outcome, adjusting for confounding. Error in covariate measurements, if ignored, results in biased estimates of parameters representing the associations of interest. Studies with variables measured with error can be considered as studies in which the true variable is missing, for either some or all study participants. We make the link between measurement error and missing data and describe methods for correcting for bias due to covariate measurement error with reference to this link, including regression calibration (conditional mean imputation), maximum likelihood and Bayesian methods, and multiple imputation. The methods are illustrated using data from the Third National Health and Nutrition Examination Survey (NHANES III) to investigate the association between the error-prone covariate systolic blood pressure and the hazard of death due to cardiovascular disease, adjusted for several other variables including those subject to missing data. We use multiple imputation and Bayesian approaches that can address both measurement error and missing data simultaneously. Example data and R code are provided in supplementary materials.
\end{abstract}

\keywords{Bayesian methods \and Maximum likelihood \and Measurement error \and Multiple imputation \and Regression calibration}

\section{Introduction}

Studies with variables measured with error can be considered as studies in which the true variable is missing, for either some or all study participants. This article focuses on measurement error in covariates in regression analyses in which the aim is to estimate the association between one or more covariates and an outcome, adjusting for confounding. Error in covariate measurements, if ignored, results in biased estimates of parameters representing the associations of interest. We make the link between measurement error and missing data and describe methods for correcting this bias. By considering covariate measurement error as a missing data problem, we can take advantage of statistical methods developed for handling missing data.
Interest is assumed to be in the regression of an outcome $Y$ on two correlated covariates, $X$ and $Z$, using the generalized linear model
\begin{equation}
   g\{ E(Y|X,Z)\}=\alpha+\beta_X X+\beta_Z Z.
    \label{eq:mod.true}
\end{equation}
We shall also consider models for censored failure time outcomes. We consider a cohort of $n$ individuals and the outcome $Y$ is observed without error on the complete cohort. Covariate $X$ is measured with error, and the error prone variable is denoted $X^*$. Covariate $Z$ is measured without error for all individuals. A naive analysis uses $X^*$ in place of $X$, with $X^*$ observed for all individuals:
 \begin{equation}
    g\{E(Y|X^*,Z)\}=\alpha^*+\beta_X^* X^*+\beta_Z^* Z.
    \label{eq:mod.err}
\end{equation}
Estimates of $\beta_X^*$ and $\beta_Z^*$ are not unbiased estimates of the parameters of interest, $\beta_X$ and $\beta_Z$. The direction of the bias in these parameter estimates depends on the form of the error in $X^*$, and the relation between $X$ and $Z$. Different forms of error in $X^*$ and their impact on parameter estimates are discussed in the next section. 

To make any correction for the impact of measurement error in $X^*$ information is needed about the relationship between $X$ and $X^*$. We consider three types of sub-study within the cohort from which this information can be obtained:

\emph{Validation study:} In a validation study the true value of $X$ is observed for a subset of individuals in the cohort. A validation study permits a wider range of possible error models to be used.

\emph{Replication study:} In a replication study two or more error-prone measures of $X$ are observed for at least a subset of the cohort, and we denote the repeated measures $X_j^*$ $j=1,\ldots,J$, with all individuals having one measure $X_1^{*}$. The true value of $X$ is not observed for any individuals in the cohort. A replication study is appropriate if the $X_{j}^*$ are subject only to independent error such that $E(X_{j}^{*}|X)=X$.

\emph{Calibration study:} A calibration study is appropriate if $X^*$, assumed to be observed for all individuals in the cohort, is subject to systematic error, that is $E(X^{*}|X)\neq X$. As in a replication study, the true value of $X$ is not observed for any individuals in the cohort. In a calibration study, a second error prone measure of $X$, denoted $X^{**}$, which is subject only to mean zero independent error is observed in at least a subset of individuals. In some situations, two or more repeated measures, $X^{**}_{j}$ $j=1,\ldots,J$, may be obtained for some individuals.

In a validation, replication and calibration study a subset of the study cohort has more information on the true value $X$, with the form of this information differing in the three study types. We let $R_i=1$ for individuals with all possible measurements observed and $R_i=0$ for individuals with a minimal set of measurements. Table \ref{tab:study.type} summarises what is observed for individuals with $R_i=1$ and $R_i=0$ in the three study types assuming $J=2$.

\begin{table}
\caption{Summary of what is observed in different study types.}\label{tab:study.type}
\centering
\begin{tabular}{lll}
\hline
Study type&$R_i=1$&$R_i=0$\\
\hline
Validation study& $\{Y_i,Z_i,X_i,X_i^*\}$&$\{Y_i,Z_i,X_i^*\}$\\
Replication study&$\{Y_i,Z_i,X^*_{1i},X^*_{2i}\}$&$\{Y_i,Z_i,X^*_{1i}\}$ \\
Calibration study (i)&$\{Y_i,Z_i,X_i^*,X_i^{**}\}$&$\{Y_i,Z_i,X_i^*\}$\\
Calibration study (ii)&$\{Y_i,Z_i,X_i^*,X^{**}_{1i},X^{**}_{2i}\}$&$\{Y_i,Z_i,X_i^*\}$\\
\hline
\end{tabular}
\end{table}


In Section \ref{sec:miss.mech} we discuss missing data mechanisms and how they relate to the measurement error setting, considering different forms of error. Sections \ref{sec:rc}-\ref{sec:bayes} outline methods for measurement error correction using regression calibration, maximum likelihood estimation, and Bayesian analysis. In Section \ref{sec:MI} we discuss the extension of multiple imputation to the measurement error setting. Section \ref{sec:example} presents a comparison of the methods through an application to data from the Third National Health and Nutrition Examination Survey (NHANES), and we conclude with a discussion in Section \ref{sec:disc} 

\section{Missing data and measurement error mechanisms}
\label{sec:miss.mech}

\subsection{Sources of missingness}

There are two sources of missingness, or coarsening, in our setting. The first is due to the error in the observed covariates, and the second is the selection of only a subset of individuals to the validation, replication or calibration sub-study ($R_i=1$ in Table \ref{tab:study.type}). The form and degree of measurement error in the measures $X^*$ (and $X^{**}$ in a calibration study) determines the extent to which the true covariate values are `missing' or coarsened and hence the degree of potential bias in naive analyses ignoring measurement error. We can view the coarsening due to measurement error as a continuum: if the measurement error is small, the error prone values $X^*$ contain substantial information about $X$, and as the degree of error increases this information is reduced. Missingness due to selection to the sub-study can be considered as regular missing data. The individuals selected to the sub-study are used to learn about the extent of error in $X^*$ and hence make corrections for it. 

When there is a validation study in which $X$ is observed for a subset of individuals, we can consider ourselves to be in a standard missing data setting: the true $X$ is observed for some individuals and missing for others. Though a difference from the standard missing data setting is that we have an observed error-prone exposure measure $X^*$ for all individuals. Such variables are sometimes referred to as \emph{auxiliary variables} in the standard missing data setting. 

\subsection{Missing data and selection mechanisms}

In the missing data context, the missingness mechanism can be classified as one of three types, as proposed by Rubin. Data are described as missing completely at random (MCAR) when the probability of the true value being observed for a given individual is unrelated to the observed and unobserved data for that individual. In the context of a validation study this can be expressed as $\mathrm{Pr}(R_i=1|Y_i,X_i,Z_i,X_i^*)=\mathrm{Pr}(R_i=1)$. Data are missing at random (MAR) when the probability of the true value being observed depends on the fully observed variables and conditional on these, not on the partially observed variable $X$, which can be expressed as $\mathrm{Pr}(R_i=1|Y_i,X_i,Z_i,X_i^*)=\mathrm{Pr}(R_i=1|Y_i,Z_i,X_i^*)$. The third possibility is that data are missing not at random (MNAR), meaning that the probability of the true value being observed depends on $X$, even after conditioning on the fully observed variables, so that $\mathrm{Pr}(R_i=1|Y_i,X_i,Z_i,X_i^*)\neq \mathrm{Pr}(R_i=1|Y_i,Z_i,X_i^*)$. We refer to Rubin \cite{Rubin:1987}, Carpenter and Kenward \cite{Carpenter:2013} and Seaman et al. \cite{Seaman:2013} for discussion and definition of missing data mechanisms. In a replication study the missingness mechanism corresponds to the mechanism that determines which subset of subjects has a replicate measure taken, and in a calibration study to the mechanism that determines which subset of subjects has the superior measurement. The MAR assumption therefore corresponds to $\mathrm{Pr}(R_i=1|Y_i,X_{1i}^*,Z_i,X_{2i}^*)= \mathrm{Pr}(R_i=1|Y_i,X_{1i}^*,Z_i)$ in a replication study and to $\mathrm{Pr}(R_i=1|Y_i,X_{i}^*,Z_i,X_{i}^{**})= \mathrm{Pr}(R_i=1|Y_i,X_{i}^*,Z_i)$ in a calibration study.

Within the context of regular missing data, the naive analysis approach is a `complete case' analysis, in which the regression is based on the subset of individuals with no missing data. A major concern is that the probability that a variable is missing is conditionally or unconditionally related to the underlying value of that variable (MAR or MNAR). If this is the case, a complete case analysis may lead to biased inferences. Even if the data are missing completely at random (MCAR) it is usually inefficient to restrict the analysis to the complete case subset, because for individuals with incomplete data, there is often some information about the parameters of interest contained in the incomplete cases' data.

In the case of measurement error, the selection mechanism is typically known because the subset with $R_i=1$ is chosen by design. The subset is almost always chosen completely at random, although alternative designs have been proposed, for example where those with an extreme first measurement are selected to have a second measurement \cite{berglund2007correction}. When, as is usually the case, the selection is completely at random, the missingness process can be ignored for the purposes of inference. This feature renders this aspect of the problem much easier than in the usual case of missing data. If the subset $R_i=1$ arises in a non-random fashion, attention should be paid to the missing data mechanism. We primarily focus on the MCAR setting. 

\subsection{Measurement error models}

In the case where a validation study is available, unbiased estimates of the parameters in model \ref{eq:mod.true} could be obtained by fitting the model within the subset of individuals with $X$ observed ($R=1$). However, this would be highly inefficient because the size of the validation sample is usually small and methods that make use of all the data should be used. This is not even an option in a replication or calibration study were $X$ is not observed for anyone, and hence other approaches are needed to obtain unbiased estimates of the parameters of interest. For most measurement error correction methods we need to understand the form and degree of the measurement error. In subsequent sections we will discuss several methods with reference to common forms of error.

The simplest form of error is classical random error, such that the observed value is the true value plus independent random error:
\begin{equation}
X_i^*=X_i+U_i.
\label{eq:err.classical}
\end{equation}
where $E(U_{i}|X_{i})=0$ and they have constant variance $\sigma^{2}_{U}$, and are independent of $X$, $Y$ and $Z$. An extension includes systematic error components as well as the random component:
\begin{equation}
X_i^*=\theta_0+\theta_1 X_i+U_i.
\label{eq:err.sys}
\end{equation}
An assumption often made is that measurement errors are \emph{non-differential}, meaning that the errors and outcome $Y$ are conditionally independent given $X$ and $Z$. A consequence of this assumption is that $E(Y|X,Z,X^*)=E(Y|X,Z)$.

There are many other types of measurement error. These include multiplicative error, in which $X^{*}$ is equal to $X$ multiplied by a random component, such that the classical error model may hold on the log scale. Some error-prone measures are subject to a threshold effect or to an excess number of zero measures. Under Berkson error the error-prone measures $X^{*}$ are less variable than the true values $X$, with $X=X^{*}+U$, $E(U|X^{*})=0$, which raises specific issues for analysis.

In a validation study, the form of the measurement error can be investigated directly because the true values are known for some individuals alongside the error prone values. In contrast, in a replication or calibration study, it is necessary to make assumptions about the form of the error in order to make progress in correcting for it.

\section{Regression calibration and conditional mean imputation}
\label{sec:rc}

Regression calibration is the most commonly applied method for measurement error correction \cite{Armstrong:1985,Shaw:2018}. It involves replacing $X$ in the regression model of interest (\ref{eq:mod.true}) with its expectation conditional on the error-prone measure and the fully observed covariates $E(X|X^*,Z)$:
\begin{equation}
    g\{E(Y|X^*,Z)\}=\alpha+\beta_X E(X|X^*,Z)+\beta_Z Z.
    \label{eq:mod.rc}
\end{equation}
In the case of a linear regression model this is justified by the result that
\begin{eqnarray*}
E(Y|X^{*},Z) &=& E\{E(Y|X,Z,X^*)|X^*,Z\} \\
&=&E\{E(Y|X,Z)|X^*,Z\} \\
&=&\alpha+\beta_X E(X|X^*,Z)+\beta_Z Z
\end{eqnarray*}
which makes the assumption that the measurement error is non-differential ($E(Y|X,Z,X^*)=E(Y|X,Z)$). Regression calibration therefore relies on the assumption of non-differential error. The justification of this method for non-linear outcome models, for example logistic regression, uses approximations \cite{Rosner:1989,Rosner:1992}. Regression calibration also extends approximately to survival studies \cite{Hughes:1993}.

To apply regression calibration requires estimation of the expectation $E(X|X^*,Z)$. In a validation study the expectation $E(X|X^*,Z)$ can be estimated directly, using a regression of $X$ on $X^*$ and $Z$ using the subset of individuals with $R=1$. For example, assuming this conditional mean is linear in $X^{*}$ and $Z$,
\begin{equation}
    X=\gamma_0+\gamma_X^* X^*+\gamma_Z Z+\epsilon,
    \label{eq:rc}
\end{equation}
giving $\hat{E}(X|X^*,Z)=\hat{\gamma}_0+\hat{\gamma}_X^* X^*+\hat{\gamma}_Z Z$. It is inefficient to use $E(X|X^*,Z)$ when $X$ is actually observed, and so in this case $X$ is used when it is observed ($R=1$). 

In a replication study, and when $X^*_1$ and $X^*_2$ are assumed to be have classical error as defined in (\ref{eq:err.classical}), the unobserved $X$ is replaced by $E(X|X^{*}_{1},Z)$ (or rather an estimate of this). This can be estimated from a linear regression of $X^*_2$ on $X^*_1$ and $Z$ in the $R=1$ group. A more efficient approach is to replace $X$ in the outcome model by $E(X|X^{*}_{1},Z)$ for those with $R=0$ and $E(X|X^{*}_{1},X^{*}_{2},Z)$ for those with $R=1$. These expectations can be in turn be estimated efficiently by fitting a random intercepts model to the data \cite{bartlett2009linear}.

In a calibration study, the expectation $E(X|X^*,Z)$ can be estimated from a linear regression of $X^{**}$ on $X^*$ and $Z$ in the $R=1$ group, under the assumption that $X^{**}$ follows the classical error model and $X^{*}$ has systematic error of the form in (\ref{eq:err.sys}). As in a replication study, more efficient use of the data could be made by making additional distributional assumptions. Only one measure $X^{**}$ is needed in a subset of individuals to perform regression calibration in a calibration study; other methods, as we will discuss below, require more than one measure $X^{**}$ in a subset of individuals. 

Regression calibration can be viewed as a `conditional mean imputation' approach. For individuals with the true value of $X$ missing, which is \emph{all} individuals in a replication or calibration study, a value is imputed as the estimated conditional mean given $Z$ and the error-prone measurements. 
In Section \ref{sec:MI} we contrast this single imputation approach with a multiple imputation approach. 

After using $\hat{E}(X|X^*,Z)$ in place of $X$ in the outcome model, standard errors can be obtained using standard methods. However, these do not take into account the uncertainty in the estimation of $\hat{E}(X|X^*,Z)$, that is the uncertainty in the estimates of the parameters in the model used to estimate $E(X|X^*,Z)$. In the above example given for a validation study (\ref{eq:rc}), this is the parameters $\gamma$. If this uncertainty is ignored, the confidence intervals for estimate of the parameters in (\ref{eq:mod.rc}) will be too narrow and therefore have too-low coverage. This can be resolved in one of two ways. In simple settings with a small number of covariates, such as that focused on in this article, a `delta method' approximation can be made to obtain correct estimated standard error \cite{Rosner:1989}. More generally, bootstrapping can be used, in which the regression calibration model (e.g. (\ref{eq:rc})) and the outcome model \ref{eq:mod.rc} are both fitted within each bootstrap sample. 

In more recent work, some authors have extended the use of regression calibration to include additional variables, $V$ say, in the set of predictors so that the focus is on estimating $E(X|X^*,Z,V)$ \cite{Freedman:2011,Prentice:2009}. The $V$ should be predictors of $X$ but not independently associated with the outcome. This approach has been applied in nutritional epidemiology, where $X$ represents a true long term average intake of a given nutrient, $X^*$ is a self-reported measure and $V$ denotes a biomarker measure. Thinking of the measurement error in the missing data context, the additional variables $V$ take the role of auxiliary variables. 


\section{Maximum likelihood}
\label{sec:ML}

In this section we briefly review the likelihood approach to measurement error correction. We refer to Carroll et al. \cite{Carroll:2006} for a detailed overview. In the likelihood approach a joint model for the full data vector must first be specified, with parameter $\omega$. The model is almost always specified conditional on the error-free covariates $Z_{i}$, to avoid having to model their distribution. In a validation study contributions to the likelihood function are thus of the form $p(Y_i,X_i, X^*_i,R_i=1|Z_i;\omega)$ and $p(Y_i,X^*_i,R_i=0|Z_i;\omega)$, with corresponding expressions in the cases of replication or calibration studies. When the selection into the substudy is MCAR or MAR, there is no need to model the missingness indicator $R_{i}$.

The observed data likelihood involves integrating over the unobserved $X$ in those portions of the study sample in which $X$ is unobserved. The likelihood in the setting of a validation study is
\begin{equation}
    L_V= \prod_{i:R_i=1}p(Y_i,X_i,X_i^{*}|Z_i;\omega)\prod_{i:R_i=0}\int p(Y_i,X_i^*,X|Z_i;\omega) dX
\end{equation}
If the measurement error is assumed to be non-differential, so that $p(Y_{i}|X_{i},Z_{i},X^{*}_{i})=p(Y_{i}|X_{i},Z_{i})$, then the likelihood can be factorised as
\begin{equation}
\begin{split}
    L_V&=  \prod_{i:R_i=1}p(Y_i|X_i,Z_i;\beta)p(X_i^*|X_i,Z_i;\theta) p(X_i|Z_i;\gamma)\\
    &\times \prod_{i:R_i=0}\int p(Y_i|X,Z_i;\beta) p(X_i^*|X,Z_i;\theta) p(X|Z_i;\gamma) dX .
    \end{split}
\end{equation}
In the above, we have used $\beta$ to denote the parameters of the outcome model, $\theta$ the parameters of the measurement error model ($\theta=\sigma^{2}_{U}$ in the case of classical error), and $\gamma$ the parameters describing the distribution of $X$ given $Z$.

Assuming again that error is non-differential, and suppressing dependence on parameters in the notation, in a replication study the observed data likelihood is
\begin{equation}
\begin{split}
    L_R&=\prod_{R_i=1}\int p(Y_i|X,Z_i) p(X_{i1}^*|X,Z_i) p(X_{i2}^*|X,Z_i) p(X|Z_i) dX \\
    &\times \prod_{R_i=0}\int p(Y_i|X,Z_i) p(X_{i1}^*|X,Z_i) p(X|Z_i) dX
\end{split}
\label{eq:ML.rep}
\end{equation}
The likelihood in the case of a calibration study is the same as that of a replication study, except the model and corresponding likelihood component for the first measurement must account for the assumed systematic measurement error.

The integrals in the observed data likelihoods over the unobserved $X$ are generally intractable except for certain special cases. One such special case is where the outcome is continuous and the full data vector is assumed to be multivariate normal, for which in the case of a replication study maximum likelihood estimates can be obtained by fitting an appropriate linear mixed model (\cite{bartlett2009linear}). More generally, the integrals must be approximated using quadrature or Monte-Carlo methods. The observed data likelihood (or an estimate of it) can then be maximised using gradient type methods. In Stata the gllamm command implements Newton-Raphson with adaptive quadature (\cite{rabe2004maximum}), and the same can be readily implemented in SAS using PROC NLMIXED. In R and Stata the merlin package provides a flexible approach to this problem \cite{Crowther:2018}


A natural concern with using fully parametric maximum likelihood is the robustness of inferences to misspecification of various parts of the joint model. To try and mitigate such concerns, a number of authors (\cite{hu1998,schafer2001,rabe2004maximum}) have proposed and investigated a semiparametric likelihood approach which avoids specification of a parametric model for $X$ (or the residuals in the $X|Z$ model).

\section{Bayesian approach}
\label{sec:bayes}

In the Bayesian approach a joint model is chosen as described in the preceding section. To this, prior distributions are added for the model parameters, and inference is based on the implied posterior distributions. Assuming \textit{a priori} independence between the parameters of the sub-models, we can do this by specifying separate priors for each sub-model parameter, i.e. $\beta$, $\delta$, $\gamma$. Bayesian inference is then based on the posterior distribution of the model parameters, which is proportional to the product of the prior and the observed data likelihood. Usually primary interest is in the parameters $\beta$ indexing the outcome model, in which case we would focus on the marginal posterior of $\beta$.

Analogous to the situation with maximum likelihood, the posterior distribution(s) are typically not available in closed form, and so we must resort to numerical methods to obtain estimates of them. In recent decades the most popular approach to this has been Markov Chain Monte Carlo (MCMC) which use Gibbs and/or Metropolis-Hastings sampling. This can be performed readily in software such as BUGS or JAGS.

Several authors have discussed Bayesian analysis for measurement error correction, including Richardson and Gilks \cite{Richardson:1993,Richardson:1993b} and Gustafson \cite{Gustafson:2003}.

In the last decade there have been a number of developments in terms of improved computational methods for obtaining Bayesian inferences when the posterior is analytically intractable, with direct relevance to measurement error correction. One is the development of Hamiltonian Monte-Carlo (HMC) methods, and its implementation in the software Stan \cite{carpenter2017stan}. HMC samplers usually converge to their stationary distribution faster and require less iterations for inference because the autocorrelations within chains tend to be much lower.

A second is inference based on the integrated Laplace approximation (INLA) approach \cite{Muff2015}. The INLA approach numerically approximates the required posteriors by using numerical integration together with Laplace approximations used to approximate the integrands. INLA has been shown to provide accurate and computationally fast estimates of the posteriors for so called latent Gaussian models. The latter includes GLMs with covariates subject to normal measurement errors.

Aside from any philosophical considerations, a practical strength of the Bayesian approach in general, and in the context of measurement error correction in particular, is the ability to bring external knowledge or information in to the analysis, via the specification of the priors. 

\section{Multiple imputation for measurement error correction}
\label{sec:MI}

Multiple imputation (MI) has become arguably the most popular principled approach for the handling of missing data in practice.  It involves generation of multiple imputations for each missing value, resulting in multiple completed datasets. As originally formulated, imputations are generated as independent draws from the posterior distribution of the missing data from an appropriate Bayesian model. The substantive model of interest \ref{eq:mod.true} is then fitted to each completed dataset. Estimates and standard errors from each analysis are then combined as we describe below. 

MI can be transferred directly from the traditional missing data setting to the measurement error setting when there is a validation study in which the true exposure $X$ is observed for a subset of individuals. There are, however, two ways in which this measurement error setting differs from the usual missing data setting. First, the proportion of individuals with $X$ unobserved will typically be smaller than the proportion of individuals with fully observed data in the missing data context. Second, all individuals have the error prone value $X^*$ observed. This takes the role of an auxiliary variable in the missing data context. MI can also be used for measurement error correction in the context of a replicates study or a calibration study, though the steps required to obtain the imputations are different from in the missing data context, and software for MI for missing data mostly cannot handle this situation.

The use of MI as a method for measurement error correction was probably first described by Brownstone and Valletta \cite{Brownstone:1996} in an economics context. They wished to correct for the impact of non-random error in measures of individual earnings, when used as a dependent variable in a regression, and had access to validation data. Cole et al. \cite{Cole:2006}, Freedman et al. \cite{Freedman:2008} and Keogh and White \cite{Keogh:2014} focused on MI in covariates in the epidemiological context, and their work is discussed further below. More recently, MI for measurement error correction has been discussed in the sociological literature by Blackwell et al. \cite{Blackwell:2017,Blackwell:2017.2}, who viewed measurement error as partially missing data, and in turn viewed missing data as an extreme form of measurement error. Their approach accommodates both measurement error and missing values, but is restricted by requiring a multivariate normal distribution for the variables with missing values and measurement error. 

In this section, we provide an overview of the use of MI for measurement error correction, considering validation, replicate and calibration studies. 

\subsection{General procedure}
\label{sec:gen.mi}

The general procedure is the same across settings. The aim is to draw values of the true exposure $X$ from its distribution conditional on the error prone measure $X^{*}$, the outcome $Y$, and any covariates $Z$. Let $p(X|X^*,Z,Y;\theta)$ denote a model for this distribution, parameterized by $\theta$. MI consists of the following steps:
\begin{enumerate}
    \item Fit the model $p(X|X^*,Z,Y;\theta)$ by maximum likelihood to obtain estimates $\hat{\theta}$ and a corresponding variance-covariance matrix $\Sigma_{\theta}$. Then for $m=1,\ldots,M$:
    \item Obtain draws $\theta^{m}$ of the parameters from their approximate joint posterior distribution.
    \item For each individual $i$ with $X$ unobserved, obtain a draw of $X_{i}^{m}$ from $p(X_{i}|X_{i}^*,Z_{i},Y_{i};\theta^{(m)})$. This gives an imputed data set in which the true exposure is available for all individuals. 
    \item Fit the substantive model to obtain estimates $\hat{\mathbf{\beta}}^{(m)}=(\hat{\alpha}^{(m)},\hat{\beta}_{X}^{(m)},\hat{\beta}_{Z}^{(m)})^{T}$ and their estimated variance-covariance matrix $\Sigma^{(m)}$.
\end{enumerate}
The estimates $\hat{\mathbf{\beta}}^{(m)}$ and $\Sigma^{(m)}$ ($m=1,\ldots,M$) are then combined using Rubin's Rules \cite{Rubin:1987} to give an overall estimate of $\beta$ and corresponding variance-covariance matrix. 

According to Rubin's Rules the pooled estimate is 
\begin{eqnarray*}
\hat{\mathbf{\beta}}= M^{-1} \sum^{M}_{m=1} \hat{\mathbf{\beta}}^{(m)}
\end{eqnarray*}
and its estimated variance-covariance matrix is 
\begin{eqnarray*}
\mathrm{var} (\hat{\mathbf{\beta}})= \mathbf{W}+(1+M^{-1})\mathbf{B}
\end{eqnarray*}
where $\mathbf{W}$ is the within-imputation variance and $\mathbf{B}$ the between-imputation variance. Rubin's rules should be applied on a scale for which the full data estimator is as close as possible to normal. 

Although in principle any model $p(X|X^*,Z,Y;\theta)$ could be used to perform the imputation, potentially serious bias could arise in the estimates of $\mathbf{\beta}$ and their variance-covariance matrix if the imputation model is mis-specified. In particular, the imputation model should be compatible with the substantive model that is used for the analysis, i.e. model (\ref{eq:mod.true}), which is assumed to be correctly specified \cite{Meng:1994}. When the substantive model is a normal linear regression with linear effects of $X$, a conditional normal model for $p(X|X^*,Z,Y;\theta)$ meets the compatibility requirement. This also holds when the substantive model is a logistic regression. However, if the substantive model is a more complex model, such as a Cox regression, then standard parametric imputation models are not strictly compatible \cite{Bartlett:2015}. Issues of compatibility also arise in general if the substantive model includes non-linear terms in $X$, including interaction terms. 

\subsection{Validation study}

As noted above, MI can be transferred directly from the traditional missing data setting to the measurement error setting when there is a validation study, with the error-prone variable $X^*$ treated as an auxiliary variable. MI for measurement error correction in covariates was described by Cole et al. \cite{Cole:2006}, who focused on a validation study, though this was only made explicit in a commentary by White \cite{White:2006}. Cole et al. \cite{Cole:2006} describe their method in the context of a single misclassified binary variable and where the substantive model is Cox regression. They suggested an imputation model which is a logistic regression of $X$ on $X^{*}$, $D$ (the binary event indicator) and $\log T$ (the smaller of the event and censoring time). This model can be fitted within the validation study. The remainder of their MI procedure is as described above. Subsequent work in the missing data context showed that an approximately compatible imputation model would be a logistic regression of $X$ on $X^{*}$, $D$ and $\hat{H}(T)$, the Nelson-Aalen estimate of the cumulative hazard \cite{White:2009}. For continuous $X$, the logistic regression is replaced by a linear regression. The substantive model compatible method of Bartlett et al. \cite{Bartlett:2015} can also be used directly. 

MI using standard software in the validation setting can be used to impute $X$ using a model for $p(X|X^{*},Z,Y)$, which allows for the measurement error to be differential. To utilise the non-differential error assumption the substantive model compatible approach can be adopted, whereby $X^{*}$ is used as an auxiliary variable to impute $X$ but is not included in the outcome/substantive model.

\subsection{Replicates and calibration studies}
\label{sec:MI.rep}

MI for measurement error correction in the context of a replicates study was described by Keogh and White \cite{Keogh:2014}. In a replicates study we do not observe the true exposure $X$ for any individuals, and so the imputation model $p(X|X^*,Z,Y;\theta)$ must be fitted indirectly. This requires repeated measures(e.g. $X^*_1,X^*_2$) and some additional distributional assumptions.  We assume a multivariate normal distribution for $(X,X_{1}^{*},X_{2}^{*})$ conditional on $Z$ and $Y$. This enables us to derive the implied model $p(X|X_{1}^{*},X_{2}^{*},Y,Z;\theta)$, required for step 1 of the general imputation procedure. Repeated measures are needed on a subset of individuals to identify the variance of the errors $U_i$ in the classical error model $X^*_{ij}=X+U_{ij}$ ($j=1,2)$. In step 3, for individuals in the sub-study with the replicate measure $X^*_2$ observed $(R_i=1)$, a draw $X_i^{(m)}$ is taken from $p(X|X_{1}^{*},X_{2}^{*},Y,Z;\theta^{(m)})$, while for individuals with only $X^*_1$ observed $(R_i=0)$, a draw is taken from $p(X|X_{1}^{*},Y,Z;\theta^{(m)})$. 

Under the classical error model for $X^*_1$ and $X^*_2$ and the multivariate distribution for $(X,X_{1}^{*},X_{2}^{*})$ conditional on $Z$ and $Y$ it can be shown that $X|X_{1}^{*},X_{2}^{*},Z,Y$ is normally distributed with mean and variance:
\begin{equation}
    E(X|X_{1}^{*},X_{2}^{*},Z,Y)=E(X|Y,Z)+\left(X_{1}^{*}+X_{2}^{*}-2E(X|Y,Z)\right)\frac{2\mathrm{var}(X|Y,Z)}{\mathrm{var}(X|Y,Z)+\sigma^{2}_U}
\end{equation}
\begin{equation}
    \mathrm{var}(X|X_{1}^{*},X_{2}^{*},Z,Y)=\frac{\mathrm{var}(X|Y,Z)}{2\mathrm{var}(X|Y,Z)+\sigma^{2}_U}
\end{equation}
Similarly it can be shown that $X|X_{1}^{*},Z,Y$ is normally distributed with mean and variance:
\begin{equation}
    E(X|X_{1}^{*},Z,Y)=E(X|Y,Z)+\left(X_{1}^{*}-E(X|Y,Z)\right)\frac{\mathrm{var}(X|Y,Z)}{\mathrm{var}(X|Y,Z)+\sigma^{2}_U}
\end{equation}
\begin{equation}
    \mathrm{var}(X|X_{1}^{*},Z,Y)=\frac{\mathrm{var}(X|Y,Z)}{\mathrm{var}(X|Y,Z)+\sigma^{2}_U}
\end{equation}

After the paper of Cole et al. \cite{Cole:2006}, the idea of using MI for measurement error correction was taken up again in 2008 by Freedman et al. \cite{Freedman:2008}, though they referred to their method as \textit{stochastic imputation}. Their focus was on the calibration substudy setting in which the error prone measure $X^*$ is subject to systematic error, but a second measure, $X^{**}$, which is subject only to classical error ($X^{**}_i=X_i+U_i$) is available for a subset of individuals. In fact, two measures of the second type, $X^{**}_1,X^{**}_2$, are needed for at least some individuals. Unlike Cole et al. \cite{Cole:2006}, who focused on a binary mismeasured covariate, Freedman et al. \cite{Freedman:2008} considered a continuous mismeasured covariate, alongside a continuous or binary outcome. The model $p(X|X^{*},X_{1}^{**},X_{2}^{**},Y,Z;\theta)$ can be derived by assuming that $(X,X^*,X_{1}^{**},X_{2}^{**})$ have a multivariate normal distribution conditional on $Z$ and $Y$, as in the replicates study setting. 

The above results rely on the assumption of a multivariate normal distribution for $(X,X_{1}^{*},X_{2}^{*})$ conditional on $Z$ and $Y$.  When the substantive model is a linear regression of $Y$ on $X$ and $Z$, this is straightforward to justify. However, for other types of outcome model, for example logistic or time-to-event models, the distribution of $(X,X_{1}^{*},X_{2}^{*})$ conditional on $Z$ and $Y$ no longer takes on a standard form. The substantive model compatible method of bartlett et al. \cite{Bartlett:2015}, which was designed to address this problem in the missing data context, was extended to the setting of measurement error correction with replicates study by Gray et al. \cite{Gray:2018}, and is implemented in the R package \mbox{SMCFCS} \cite{smcfcs}. We give a brief overview of the steps used in this procedure:
\begin{enumerate}
\item Draw a candidate value for each individual for whom the true exposure $X$ is missing from a `proposal distribution' $p(X|X^{*},Z)$. While this step is straightforward for a validation study, in a replicates study, it requires indirect estimation of
the parameters of a model for the proposal distribution $p(X|X^{*},Z;\gamma,\theta)$, derived from the measurement error model and a model for $X$ given $Z$. Parameter values $\gamma^*,\theta^*$ are first drawn from their approximate joint posterior and proposed values $X^c$ are then drawn from the distribution $p(X|X^{*},Z;\gamma^*,\theta^*)$.
\item A rejection rule is used to determine whether the proposed value $X^c$ for a given individual is accepted as a value from the target posterior distribution $p(X|X_{1}^{*},X_{2}^{*},Y,Z;\beta,\gamma,\theta)$. The rejection rule is derived based on the form of the substantive model and uses draws of the substantive model parameters, $\beta$. Steps 1 and 2 are repeated until a
candidate value of $X$ is accepted for every individual. The algorithm is then repeated iteratively until the imputed $X$ values have converged to a stationary distribution. The last cycle of imputed values is retained to create an imputed data set.
\item Steps 1 and 2 are repeated to create several imputed data sets. The outcome model of interest is then fitted to each imputed data set and Rubin's rules are applied as described above. 
\end{enumerate}

The proposal distribution used in step 1 is a posterior distribution and as such is based on both a prior distribution and the likelihood. Following Gustafson \cite{Gustafson:2003} proper inverse-Gamma priors are recommended for the variances in formulating the proposal distribution and the priors for regression coefficients are independent normal distributions with zero means and infinite variance. The substantive model compatible approach could also be extended to the setting of a calibration study, but this has not been considered to date.

\section{Comparison of methods}

We have described four approaches to measurement error correction. In this section we compare some of their properties and summarise results from authors who have compared the methods. 

Regression calibration, a mean-imputation approach, is attractive in its simplicity, and this had led to it becoming the most commonly applied measurement error correction approach \cite{Shaw:2018}. However, it is restricted to non-differential error. It also has a number of additional drawbacks. These include that it requires an approximation when the substantive model is not a linear regression. Extra steps, such as bootstrapping, are required to obtain correct standard errors for estimated parameters in the substantive model, which are not obtained directly in regression calibration. Further, the error prone variable must be entered in the substantive model as a linear term. Extensions to including transformations $f(X)$ in the substantive model are not in general straightforward as they would require derivation of $E(f(X)|X^{*},Z)$.

The maximum likelihood approach is attractive in that the resulting estimators are consistent, asymptotically normal and efficient. However, in the measurement error context, a maximum likelihood analysis typically requires numerical integration, and, while some software is available for doing this, it is not straightforward for all outcome models, e.g. time-to-event models. A number of authors have compared regression calibration to maximum likelihood (\cite{schafer1996likelihood,messer2008maximum,bartlett2009linear}). They show that in certain settings, e.g. large measurement error and/or strong association between $X$ and $Y$, maximum likelihood estimators can have superior efficiency to RC. Moreover, since measurement error corrected estimators typically have skewed distributions, inferences based on likelihood ratio tests and intervals typically perform better than Wald type tests and intervals. Bartlett et al. \cite{bartlett2009linear} also showed that the MLE when the outcome model is normal linear regression and $X$ is assumed normal is consistent regardless of whether the latter assumption is true.

Relative to a maximum likelihood based analysis, a Bayesian analysis requires specification of priors for the model parameters. This can be very useful when only external information is available regarding the measurement error model parameters, which can be encoded into the priors. Even when information about these is available within the study, as demonstrated by Bartlett and Keogh \cite{Bartlett:2018}, mildly informative priors can be used to stabilise inferences in settings where non-Bayesian methods such as RC or ML may be highly variable. They showed that in the internal replication study setting, Bayes estimators can have lower bias and less variability than RC, and Bayes 95\% credible intervals performed well with regards frequentist coverage.

MI can be viewed as an approximation to a full Bayesian analysis \cite{Carpenter:2013}. From the Bayesian perspective, application of MI and Rubin’s rules can be viewed as a particular route to performing a Bayesian analysis, in which one effectively assumes that the posterior distributions for the parameters are normally distributed. Like a maximum likelihood or fully Bayesian approach, MI can accommodate differential measurement error. MI is an attractive option for measurement error correction in the context of a validation studies, as the analyst can then take advantage of the many packages that have been developed in different software for the standard missing data setting. MI is less attractive in the context of a replicates or calibration study because of the lack of software. The simpler multivariate normal approach described above for implementation could be applied without specialized software. However, for a more general situation, such as the inclusions of non-linear terms for $X$ in the substantive model, the researcher would need to derive the appropriate form for the imputation model, e.g. $p(X|X^*_1,X^*_2,Z,Y;\theta)$ in a replicates study, raising complex issues of compatibility. The substantive model compatible approach outlined for replicates study resolves this problem, but it has been found to be computationally intensive and may be inefficient relative to a direct Bayesian analysis \cite{Gray:2018}. Lastly, the use of Rubin's rules to obtain pooled estimates assumes normality of the estimators. In the measurement error context the estimators are not in general normally distributed, which calls into question the validity of Rubin's rules. For these reasons, Bartlett ad Keogh \cite{Bartlett:2018} argued that a direct Bayesian approach may in many cases be preferable to use of MI.

Freedman et al. \cite{Freedman:2008} compared the MI approach with regression calibration and another imputation-based approach called `moment-adjusted imputation'. They found that regression calibration can be more efficient than MI when the measurement error is non-differential, but importantly their implementation of MI did not exploit the non-differential error assumption. They reported that in further simulations an implementation of MI that did utilise the non-differential error assumption had similar efficiency to RC. Under differential error, regression calibration gave biased estimates while their differential MI implementation delivered approximately unbiased estimates. 
 

\section{Example: Risk factors for cardiovascular disease}
\label{sec:example}

\subsection{Data description}

To illustrate the methods outlined in this article, we consider how they can be applied to a research question using data from the Third National Health and Nutrition Examination Survey (NHANES III). These data were used by Bartlett and Keogh \cite{Bartlett:2018} in an empirical comparison of a Bayesian analysis with regression calibration. The data are freely available at \url{https://wwwn.cdc.gov/nchs/nhanes/Default.aspx} and example code is made available alongside the data set used at \url{https://github.com/ruthkeogh/meas_error_handbook}.

NHANES III was a survey conducted in the United States between 1988 and 1994 in 33,994 individuals aged two months and older. We consider a model relating known risk factors for cardiovascular disease (CVD) measured at the original survey to subsequent hazard of CVD. Mortality status at the end of 2011 is available through linkage to the US National Death Index, with cause of death classified using the ICD-10 system. We restricted to individuals aged 60 years and above at the time of the original survey. 

The error-prone variable in this example is systolic blood pressure (SBP) and the covariates $Z$ to be adjusted for are age, sex, diabetes status, and smoking status. The true SBP measure of interest, $X$, is assumed to be the underlying average SBP, which is unobserved. The first error-prone measurement, $X^{*}_{1}$, is the SBP measurement taken at the original survey and this is assumed to be subject to classical measurement error. An approximate 5\% subset of individuals was (non-randomly) selected to participate in a second examination, during which SBP was again measured. This second exam took place on average 17.5 days after the first exam. We assume this second measurement of SBP, $X^{*}_{2}$, to be an independent error-prone measurement of each individual’s underlying SBP. We are therefore in the setting of a replicates study. 

After deleting seven individuals who were missing diabetes status, data were available on 6519 individuals. Age and sex were observed for all individuals. The SBP measurement from the first examination, $X_1^*$, was available for 5033 (77.2\%) individuals, and the second SBP measurement $X_2^*$ was available in 401 (6.2\%) of individuals. Unfortunately, smoking status was only recorded in 3433 (52.3\%) of individuals. The analysis thus required handling of both the measurement error in the SBP measurements and the substantial missingness in the smoking and SBP variables.

By the end of 2011 1469 (22.5\%) had died due to CVD, 3641 had died from other causes, and 1409 were still alive. The analysis is based on a Weibull model for hazard for CVD: $h(t)=rt^{r-1}\exp\{\beta_0+\beta_1 \mathrm{SBP}+\beta_2 \mathrm{sex}+\beta_3 \mathrm{age}+\beta_4 \mathrm{smoker}+\beta_5 \mathrm{diabetes}\}$. Deaths from causes other than CVD were treated as censorings. 

\subsection{Analysis methods}

The combination of both measurement error and missing data is a challenge for the analysis. The time-to-event outcome also presents a challenge for some methods.

If using regression calibration to address the measurement error, missing data could in principle be handled using MI by first obtaining multiply imputed datasets in which smoking status and SBP at the original survey are imputed, and secondly applying regression calibration within each imputed dataset. However, to implement Rubin's rules requires estimates of the variance of estimates of substantive model parameters obtained through regression calibration, which can only be obtained through bootstrapping or by deriving analytical expressions. This approach has not been investigated and is impractical without readily available software. We therefore did not pursue the combination of regression calibration and MI further. 

With a Weibull model for the CVD hazard, a maximum likelihood analysis is challenging to implement because numerical integration is required to perform the integration over $X$ in equation \ref{eq:ML.rep}. While there exist general software solutions to this problem, as noted in Section \ref{sec:ML}, it does not extend easily to the measurement error setting according to our investigations. There are practical software difficulties in incorporating any one of a time-to-event outcome, a replicate measure being available in only a subset of participants, and missing data in addition to measurement error. We therefore did not pursue a maximum likelihood analysis for these data. 

Using a fully Bayesian analysis it is straightforward, given the flexible JAGS software, to incorporate both measurement error and missing data simultaneously, as well as the Weibull hazard model. We assumed that each individual’s true underlying SBP around the time at which the first measurement was obtained, $X^*_1$, was normally distributed conditional on smoking, sex, age, and diabetes, with $N(0,10^4)$ priors on the regression coefficients and a $Ga(0.5,0.5)$ prior on the precision parameter. We assumed the classical error model for the two SBP measurements, $X_{ij}^*=X_{i}+U_{ij}$ ($j=1,2$). The measurement errors $U_{i1}$ and $U_{i2}$ were considered normally distributed with means 0, a common variance $\sigma^{2}_{U}$ and zero correlation. The prior for $\sigma^{-2}_{U}$ was $Ga(0.5,0.5)$. To accommodate missingness in the smoking and SBP variables under a missing at random assumption, we assumed a model for the distribution of smoking, conditional on the fully observed error-free covariates sex, age, and diabetes. In our analysis, we assumed a logistic model for this conditional distribution and used with independent mean zero normal priors for the regression coefficients, each with variance $10^4$. For the shape parameter of the Weibull hazard model, $r$, we assumed an exponential prior with parameter 0.001. We used independent $N(0,10^6)$ priors for the scaled regression coefficients $-\beta_k/r$ ($k=0,\ldots,5$). We refer to Bartlett and Keogh \cite{Bartlett:2018} and Gustafson \cite{Gustafson:2003} for further discussion of suitable priors in the measurement error setting. 

MI naturally accommodates both measurement error and missing data, though we are not aware of any prior applications to this setting. The first MI approach outlined in Section \ref{sec:MI} for measurement error correction in a replicates study relies on the assumption of a multivariate normal distribution for $(X,X_{1}^{*},X_{2}^{*})$ conditional on $Z$ and $Y$. In our setting of a time-to-event, the outcome $Y$ for each individual is comprised of the earlier of their event or censoring time, $T$, and the event indicator $D$. The multivariate normality assumption is not compatible with a Weibull model for the hazard. We therefore considered the substantive model compatible approach outlined in steps 1-3 in section \ref{sec:MI.rep}. Missing data in smoking status is assumed to depend on the other covariates through a logistic model. The SMCFCS package in R was used to for multiple imputation, assuming a Cox proportional hazards outcome model \cite{smcfcs}. A Cox regression model for the hazard was considered for the direct Bayesian analysis \cite{Bartlett:2018}, however the computation was found to be prohibitively slow. 

\subsection{Results}

We first implemented a naive analysis ignoring both measurement error and missing values. This uses $X^*_1$ as the SBP measurement and is restricted to the subset of 2667 individuals (the `complete cases') with both $X^*_1$ and smoking status observed. On the same subset of individuals, i.e. ignoring missing data, measurement error correction was then performed on the same subset, using regression calibration, a direct Bayesian analysis, and MI. Lastly, the direct Bayesian analysis and MI were applied to additionally incorporate the missing data. These analyses are based on the full set of 6519 individuals. The results are shown in Table \ref{tab:nhanes}.

The results from RC, the direct Bayesian analysis, and MI in which the missing data in SBP and smoking status is ignored are very similar. Compared to the naive analysis, all three correction methods results in a larger estimated coefficient for SBP. The estimates for sex, age, smoking status and diabetes are also very similar across these three analyses, and similar to those from the naive analysis. The direct Bayesian and MI analyses that additionally incorporate the missing data give very similar results to each other for the covariates measured without error, though the coefficients for SBP are somewhat different. This may be because the substantive model compatible MI approach assumed a Cox rather than Weibull outcome model. As we would expect, the width of the confidence/credible intervals is significantly reduced relative to those from the analyses that ignore the missing data. There are some changes in the estimated coefficients which may be indicative of bias in the complete case analyses, for example the coefficients for SBP are a little larger when the missing data is accommodated. The widths of confidence/credible intervals are substantially reduced in the direct Bayesian and MI analyses that make use of all available data. 

\begin{table}
\caption{Application of measurement error correction methods to NHANES III data. Log hazard ratio estimates and 95\% confidence intervals (naive, RC and MI) or credible intervals (Bayes). The Bayesian estimates are posterior means.}\label{tab:nhanes}
\centering
\begin{tabular}{llll}
\hline
Covariate&Naive&Bayes$^{\footnotesize\mbox{c}}$&Bayes$^{\footnotesize\mbox{d}}$\\
\hline
SBP&0.085 (0.014, 0.157)&0.114 (0.015, 0.211)&0.121 (0.055, 0.186)\\
Male&0.49 (0.30, 0.67)&0.49 (0.30, 0.68)&0.47 (0.36, 0.57)\\
Age&0.88 (0.77, 0.99)&0.87 (0.75, 0.98)&1.02 (0.94, 1.09)\\
Smoker&0.26 (0.07, 0.46)&0.26 (0.07, 0.45)&0.25 (0.08, 0.42)\\
Diabetes&0.50 (0.29, 0.72)&0.50 (0.27, 0.72)&0.68 (0.55, 0.82)\\
&&&\\
\cline{2-4}
&RC$^{\footnotesize\mbox{c}}$&MI$^{\footnotesize\mbox{c}}$&MI$^{\footnotesize\mbox{d}}$\\
\cline{2-4}
SBP&0.114 (0.011, 0.222)&0.120 (0.020, 0.219)&0.104 (0.035, 0.173)\\
Male&0.49 (0.32, 0.68)&0.49 (0.30, 0.67)&0.46 (0.35, 0.56)\\
Age&0.87 (0.76, 0.99)&0.88 (0.77, 0.99)&1.04 (0.97, 1.11)\\
Smoker&0.26 (0.07, 0.45)&0.26 (0.07, 0.46)&0.26 (0.09, 0.43)\\
Diabetes&0.50 (0.28, 0.72)&0.50 (0.29, 0.72)&0.69 (0.56, 0.83)\\
\hline
\end{tabular}

\footnotesize{$^{\mbox{a}}$per 20mmHg.}\\
\footnotesize{$^{\mbox{b}}$per 10 years.}\\
\footnotesize{$^{\mbox{c}}$Without handling of missing data in SBP and smoking status.}\\
\footnotesize{$^{\mbox{d}}$With handling of missing data in SBP and smoking status.}
\end{table}

\section{Discussion}
\label{sec:disc}

In this article we have outlined how measurement error in covariates can be viewed as a type of missing or `coarsened' data. This has the advantage of enabling the analyst to exploit existing methods and potentially software developed for handling missing data. We considered three different settings: validation studies, replicates studies and calibration studies (Table \ref{tab:study.type}) in which different sources of information about error-prone variables are available. The common approach of regression calibration for measurement error correction can be viewed as a mean imputation approach. This was contrasted with likelihood based methods of maximum likelihood estimation, a direct Bayesian approach and MI, all of which are well-developed for their use in the traditional missing data setting. These methods offer greater flexibility in terms of modelling assumptions and in some settings more accurate inferences. They can also naturally incorporate missing data in addition to measurement error. As discussed in section \ref{sec:miss.mech}, the assumptions required for use of these approaches in the measurement error setting may be more plausible than they are in the traditional missing data setting, because the set of individuals on whom gold standard or replicate measures are available is often a random subset or has been selected by some other design known to the analyst. 

Likelihood based methods are particularly appropriate for measurement error correction in the setting of a validation study in which the `true' value of an error-pone variable is measured on some individuals. This is very close to a traditional missing data problem, with the special feature that an auxiliary variable $X^*$ is available on everyone, meaning that methods and software for traditional missing data can be applied directly. 

In Section \ref{sec:example} we discussed the application of measurement error correction methods to investigate the association between cardiovascular disease risk factors and the hazard of CVD death using data from NHANES III. Repeated measures of the error-prone variable of systolic blood pressure were available from a replicates substudy. The setting is representative of many studies in epidemiology, involving a time-to-event outcome and missing values in covariates as well as measurement error. This highlighted some of the challenges faced in applying measurement error correction methods in practice, notably that other features of the analysis need to be accommodated alongside the measurement error correction. While regression calibration is an attractively simple approach to measurement error correction, it does not extend easily to enable the analyst to simultaneously address missing values. A maximum likelihood approach is conceptually straightforward and accommodates both measurement error and missing values, but there is a lack of software for flexible implementation that allows the analyst to handle both of these issues at the same time, especially alongside a time-to-event outcome. A direct Bayesian analysis was the most flexible approach and by far the most straightforward to apply to accommodate both measurement error and missing data. MI, while attractive due to it's now common use in handling missing data, was not straightforward to implement for measurement error correction in a replicates study, or to combine with missing values.

We have focused on measurement error in a single covariate. In some settings we face error in multiple covariates, which may include both continuous and categorical variables. Regression calibration extends to the setting of multiple covariates measured with error, though its other limitations discussed above remain. Similarly maximum likelihood estimation extends directly to this setting in theory, but the practical implementation becomes further complicated by the need for multivariate numerical integration. A Bayesian approach remains straightforward both conceptually and from a practical standpoint when there are multiple covariates measured with error. The developments of MI for use in a replicates or calibration study do not appear to have extended to the setting multiple covariates with error and further development would be needed. 

In conclusion, we recommend a Bayesian analysis for measurement error correction due to its flexibility to handle several issues simultaneously and due to the readily available software for straightforward practical implementation. While initially seeming attractive, MI is currently impractical as a measurement error correction method in many realistic problems. There is a need for development of software to enable implementation of maximum likelihood estimation, particularly for those who may object on philosophical grounds to a Bayesian analysis.


\bibliographystyle{unsrt}

\end{document}